\documentclass[twoside,epsfig]{article}
\usepackage{fleqn,espcrc2}

\usepackage{epsfig}

\title{First Neutrino Observations from the Sudbury Neutrino Observatory}

\author{A.B. McDonald, Queen's University, Kingston, Ontario, Canada\\ For the SNO 
Collaboration$^a$} 

\begin{document}
\begin{abstract}
The first neutrino observations from the Sudbury Neutrino Observatory are 
presented from preliminary analyses. Based on energy, direction and location, 
the data in the region of interest appear to be dominated by $^{8}$B solar neutrinos, 
detected by the charged current reaction on deuterium and elastic scattering 
from electrons, with very little background. Measurements of radioactive 
backgrounds indicate that the measurement of all active neutrino types via the 
neutral current reaction on deuterium will be possible with small systematic 
uncertainties. Quantitative results for the fluxes observed with these reactions 
will be provided when further calibrations have been completed.
\vspace{1pc}
\end{abstract}

\maketitle

\section{INTRODUCTION}
This paper presents the first neutrino observations from the Sudbury Neutrino 
Observatory \cite{One} (SNO), a 1000 tonne heavy-water-based Cerenkov detector situated 
2 km underground in INCO's Creighton mine near Sudbury, Ontario, Canada. The SNO 
detector has been filled with water since May, 1999. After a commissioning 
period, the detector parameters were fixed at the start of November 1999 and 
neutrino data acquisition and associated calibrations have been taking place 
almost continuously since then. In this initial phase of the project, the 
detector is filled with pure heavy water. Neutrinos from $^{8}$B decay in the sun 
are observed from Cerenkov processes following these reactions:
\\
1. The Charged Current (CC) reaction, specific to electron neutrinos:

\begin{equation}
       d + \nu_{e} \rightarrow p + p + e^{-}  \nonumber
\end{equation}

This reaction has a Q value of -1.4 MeV and the electron energy is strongly 
correlated with the neutrino energy, providing very good sensitivity to spectral
distortions.
\\
2. Neutral Current (NC) reaction, equally sensitive to all non-sterile neutrino 
types:

\begin{equation}
        \nu_{x} + d \rightarrow n + p + \nu_{x}  \nonumber
\end{equation}

This reaction has a threshold of 2.2 MeV and is observed through the detection 
of neutrons by three different techniques in separate phases of the experiment.
\\
3. Elastic Scattering (ES) reaction: 
\begin{equation}
        \nu_{x} + e^{-} \rightarrow e^{-} + \nu_{x}  \nonumber
\end{equation}
This reaction has a substantially lower cross section than the other two and is 
predominantly sensitive to electron neutrinos; they have about six times greater 
cross-section than $\mu$ or $\tau$ neutrinos.

The reaction:

\begin{equation}
       \bar\nu_{e} + d \rightarrow n + n + e^{+}
\end{equation}
also provides a unique signature for anti-electron neutrinos from various 
possible sources.
\\
The SNO experimental plan calls for three phases of about one year each wherein 
different techniques will be employed for the detection of neutrons from the NC 
reaction. During the first phase, with pure heavy water, neutrons are observed 
through the Cerenkov light produced when neutrons are captured in deuterium, 
producing 6.25 MeV gammas. In this phase, the capture probability for such 
neutrons is about 25\% and the Cerenkov light is relatively close to the 
threshold of about 5 MeV electron energy, imposed by radioactivity in the 
detector. (Figure 1).  For the second phase,  about 2.5 tonnes of NaCl will be 
added to the heavy water and neutron detection will be enhanced through capture 
on Cl, with about 8.6 MeV gamma energy release and about 83\% capture efficiency. 
(See Figure 1). For the third phase, the salt is removed and an array of $^{3}$He-
filled proportional counters will be installed to provide direct detection of 
neutrons with a capture efficiency of about 45\%.

\section{PHYSICS OBJECTIVES}
The main physics goals for the Observatory are observations of:
\\
- Solar Neutrinos
\\
- Atmospheric Neutrinos 
\\
- Supernova Neutrinos
\\
- Cosmic Ray Muons
\\
- Anti-electron neutrinos from various processes including transformations of 
solar neutrinos or relic supernova neutrinos.
\\

For Solar Neutrinos, the combination of three detection reactions provides 
several sensitive ways to seek evidence for neutrino flavor change without 
relying on calculations of initial fluxes from solar models. The ratio of 
neutrino fluxes above a threshold, as observed by the CC and NC reactions 
provides a very sensitive way to observe transformations to active neutrinos.

\begin{figure}[ht]
\begin{center}
\mbox{\epsfig{file=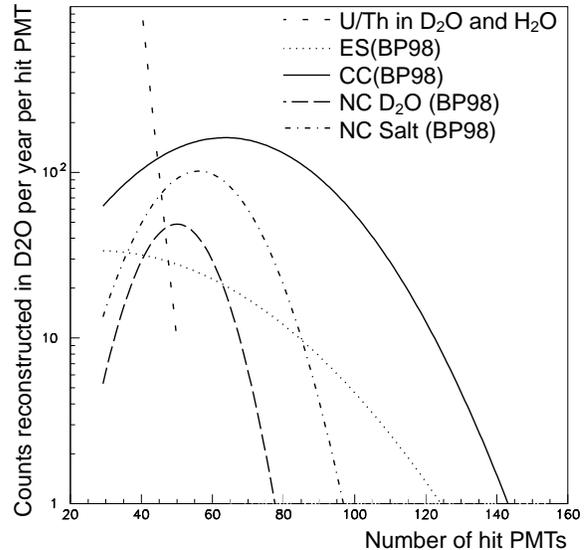,height=7cm}}
\caption{Simulations of spectra obtained from the three detection reactions 
(CC,ES,NC)for neutrino fluxes as calculated\cite{Two} by BP98. Spectra from the NC 
reaction are shown for pure heavy water and with added salt. The expected 
counting rate from U and Th radioactivity in the water is also shown.  An MeV of 
electron energy corresponds to about 9 photomultipliers (PMT's) hit.}
\end{center}
\end{figure}

The ratio of CC/NC can be observed during all three phases of operation. The 
sensitivity to the NC reaction is limited during the first phase, but there will 
be excellent sensitivity with different systematic uncertainties during the 
other  two phases. The ratio of fluxes detected by the CC and ES reaction has a 
smaller dependence on flavor change to active species through the sensitivity to 
$\mu$ and $\tau$ neutrinos in the ES cross section. 
Events from the CC and ES reactions can be distinguished through the very 
different directional response. The ES reaction is strongly peaked away from the 
Sun, whereas the CC reaction has a form of approximately $1 - 1/3$ cos$ \theta_{sun}$, 
with about a factor of two difference in rate between forward and backward 
directions relative to the Sun. The angular resolution of the detector is better 
than 25 degrees. The NC rate may be determined during the pure D$_2$O phase partly 
through a distinctive variation as a function of radius. However, the definition 
of the number of events observed with this reaction is clearly enhanced by the 
addition of salt (see Figure 1), and will be determined independently of the 
Cerenkov signals when the $^{3}$He-filled proportional counters are installed. The 
observed  spectrum for the CC reaction is a very sensitive indicator of 
distortions caused by the MSW effect\cite{Three}  because the energy of outgoing electrons 
is strongly correlated with the incoming neutrino energy and the detector energy 
resolution is better than 20\% for the range of interest.
With the relatively high statistical accuracy indicated by Figure 1, the SNO 
detector will also provide sensitive measurements of the solar neutrino flux as 
a function of zenith angle to search for MSW regeneration in the Earth. 
Correlations between flux, energy spectrum, zenith angle and time of year will 
also be studied. With the variety of reactions to be studied, the SNO detector 
can explore oscillations via the MSW effect or vacuum oscillation processes over the 
full range of  parameters consistent with previous experiments. It could provide 
clear evidence for electron neutrino flavor change, including transformations to 
either active or sterile types.

\section{DETECTOR PERFORMANCE}
The SNO detector consists of 1000 tonnes of pure D$_2$O contained within an 
acrylic vessel (12 m diameter, 5 cm thick), viewed by 9438 PMT's mounted on a 
geodesic structure 18 m in diameter, all contained within a polyurethane-coated 
barrel-shaped cavity (22m diameter by 34 m high). The cavity volume outside the 
acrylic vessel (AV) is filled with purified H$_2$O. There are 91 PMT's looking 
outward from the geodesic structure, viewing the outer H$_2$O volume.
\\
The SNO detector has been full of water since May, 1999. 
During the period until 
November, 1999, detailed commissioning tests and calibrations were performed.  
There has been no substantial problem from electrical breakdown of high voltage 
connectors submerged in light water since nitrogen was added to the degassed 
water. During the period before November, a variety of adjustments were made to 
improve the light sensitivity by about 25\% and to reduce the trigger threshold 
to about 2 MeV. Four additional PMT's were installed in the neck of the acrylic 
vessel to provide a clear indication of instrumental light emitted in this 
region, probably from static discharges of insulating materials.  $^{222}$Rn gas in 
the air above the heavy water was reduced to acceptable levels by flushing with 
boil-off gas from liquid nitrogen.
\\
As of  November, 1999, the desired detector specifications had been met, the 
detector 
parameters were frozen and production data accumulation was started, 
interspersed with a variety of calibration measurements. The detector 
performance has been very good, with more than 98.5\% of all channels 
operational; a total event rate of less than 5 Hz above a threshold of about 20 
hit PMT's; PMT individual noise rates of less than 500 Hz for a threshold of 
about 0.3 photoelectrons, providing fewer than 2 noise hits per event. 
\\
\section{CALIBRATION}

Detector calibration is being carried out with a variety of techniques and 
sources.  Electronic calibrations of pedestals, slopes and timing are performed 
regularly with pulsers. The 600,000 electronic constants are very stable. 
Optical properties of the detector have been studied using a diffusing ball, 
(Laserball) receiving light from a pulsed laser system providing wavelengths 
between 337 and 700 nm with variable intensity at repetition rates from near 0 
to 45 Hz. This source  and other calibration sources are moved within the D$_2$O 
volume using a manipulator system capable of positioning them to better than 5 
cm. Positions in the H$_2$O volume  between the D$_2$O and the PMT's are also 
accessible along vertical paths from above. A nearly mono-energetic $^{16}$N gamma 
ray source has also been deployed.

\begin{figure}[!ht]
\begin{center}
\mbox{\epsfig{file=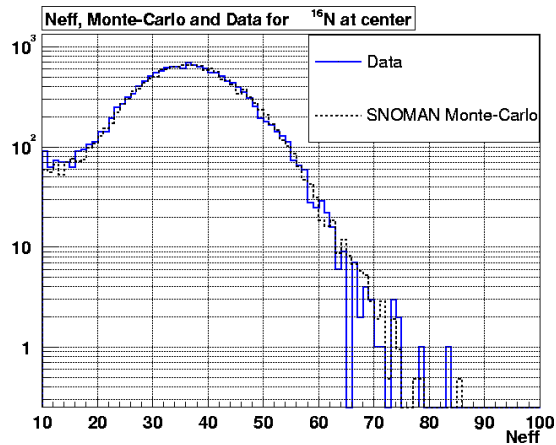,width=8cm}}
\caption{Data from the $^{16}$N source compared with Monte Carlo simulation. Neff is the number of PMTs hit by prompt light less the average number of noise hits.}
\end{center}
\end{figure}

Figure 2 shows a spectrum from the $^{16}$N source compared with a Monte Carlo 
simulation, using optical parameters extracted 
from a preliminary analysis of the laserball data. A single constant 
corresponding to the average quantum efficiency of the PMT's has been adjusted 
to match the centroid of these spectra. A further comparison of  centroids for 
over 20 other locations throughout the D$_2$O volume showed less than 2\% 
difference between the data and the simulation at any point. An acrylic-
encapsulated $^{242}$Cf fission neutron source has also been deployed to study the 
neutron response of the detector. \\
Other sources being prepared include a 19.8 MeV gamma source produced by the 
(p,t) reaction, a triggered source for the $^{232}$Th and $^{238}$U chains producing 
2.6 and 2.4 MeV gammas and a source of $^{8}$Li, emitting betas up to 13 MeV. The 
short-lived $^{16}$N and $^{8}$Li activities are produced  by a pulsed neutron 
generator located near the SNO detector and are transported via capillary tubing 
to decay chambers within the detector volume.

\section{OBSERVATIONS TO DATE}

In addition to Cerenkov light produced by neutrinos and radioactivity, there can 
be other sources of  "instrumental light" arising from parts of the detector. 
For example, it is well known that PMT's can occasionally emit light, perhaps 
through internal electrical discharges. Light from these sources has very 
different characteristics from the typical patterns observed for Cerenkov light at 
solar neutrino energies. The light from a flashing PMT shows an early trigger 
for the flashing PMT, followed by light observed across the detector, at least 
70 ns later. For SNO, six or more electronic channels surrounding the flashing 
PMT typically show pickup signals, distinguishing the events further from 
Cerenkov events.  \\
Figure 3 shows the raw spectrum of events (solid line) observed 
with the detector for a fraction of the data obtained since the start of data 
taking in November, 1999. The events are plotted against NHIT, the number of 
PMT's contained in a 400 ns second window surrounding the detector trigger (more 
than about 20 PMT's hit within a 100 ns window). NHIT is approximately 
proportional to the electron energy for a Cerenkov event, with about 9 NHIT 
corresponding to 1 MeV. Only a fraction of the data have been shown as the 
remainder are being saved for a comparison after the cuts have been fully 
defined.

\begin{figure}[!ht]
\begin{center}
\mbox{\epsfig{file=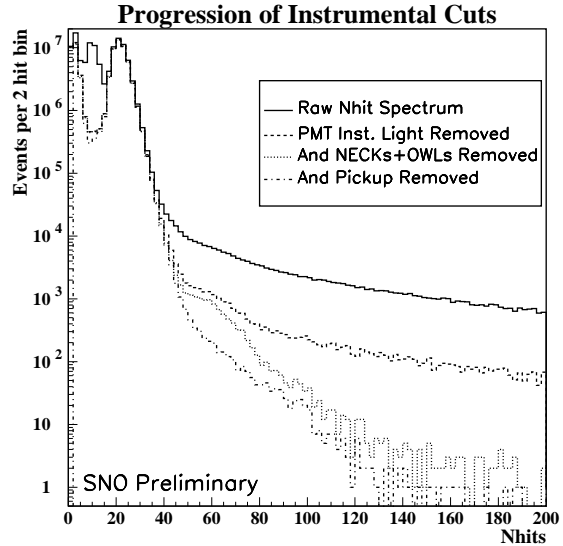,width=7cm}}
\caption{Progression of instrumental cuts.}
\end{center}
\end{figure}

The dashed line shows the residual data after cuts have been imposed to remove 
events that show characteristics matching the Flashing PMT's. The dotted line 
shows the residual data after further cuts are imposed to remove another class 
of events associated with bursts of light from the neck region of the detector. 
These events may arise from static discharges of insulating materials. Four 
additional PMT's were installed in this region in September, 1999. They clearly 
observe these events and are very insensitive to light generated in the detector 
itself, as determined from the calibration sources. The dot and dash line shows 
the residual events after the imposition of further cuts which eliminate events 
that show characteristics of pickup in the electronic systems.

Two separate groups within the SNO Collaboration developed a series of cuts to 
eliminate these instrumental light sources and their results for the residual 
spectrum were virtually identical, lending confidence in the robustness with 
which these events can be distinguished from neutrinos.
To ensure that these cuts do not remove a significant number of neutrino events, 
the fraction of signal loss was tested with the $^{16}$N source. The results are 
shown in Figure 4, indicating very low loss of signal in the region tested.
\vspace{1pc}
\begin{figure}[!ht]
\begin{center}
\mbox{\epsfig{file=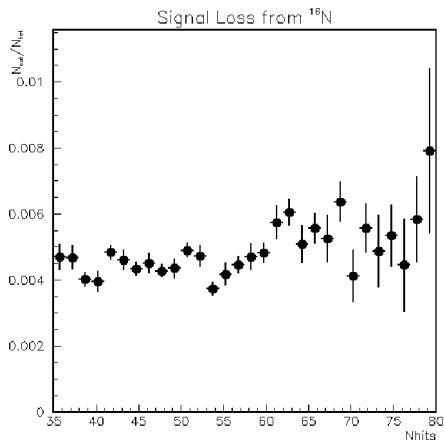,width=6cm}}
\caption{Signal loss as measured with $^{16}$N.}
\end{center}
\end{figure}

Following these cuts, algorithms based on timing and spatial information were 
used to reconstruct the position and direction of the events. 
Figure 5 shows the resulting spectrum for a large fiducial volume.

\begin{figure}[!ht]
\begin{center}
\mbox{\epsfig{file=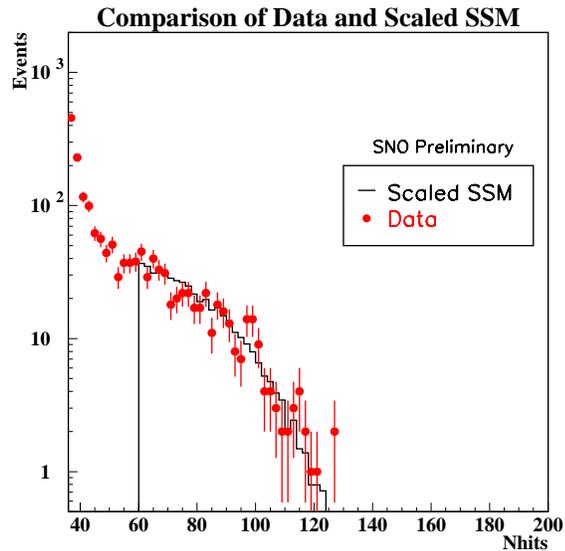,width=7cm}}
\caption{Distribution of events versus number of hit PMTs.}
\end{center}
\end{figure}

Superimposed on the data is the simulated spectrum for the CC reaction in Figure 
1, scaled to the data. As the calibrations are not yet complete, the SNO 
collaboration has chosen not to quote a number for the flux of electron 
neutrinos measured by the CC reaction on deuterium. However, it should be 
apparent from the figure that the spectrum is well defined so that an accurate 
measurement will be obtained when further calibrations have been completed.

\begin{figure}[!ht]
\begin{center}
\mbox{\epsfig{file=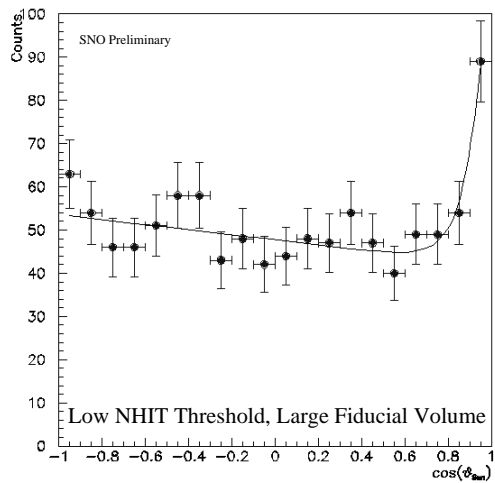,width=6cm}}
\caption{Distribution of events versus cos$ \theta_{sun}$.}
\end{center}
\end{figure}

Figure 6 shows events as a function of the direction to the sun for a lower 
energy threshold and a larger fiducial volume. Even with somewhat more 
radioactive background included by these parameter choices, the peak at cos$ \theta_{sun} = 1$ from the ES reaction is apparent.

\begin{figure}[!ht]
\begin{center}
\mbox{\epsfig{file=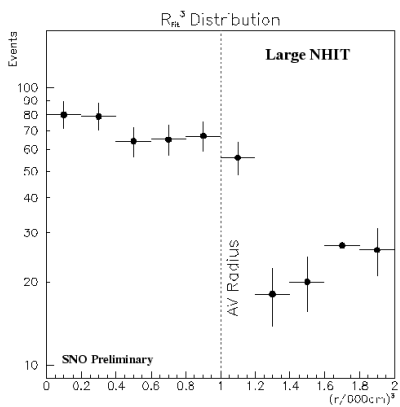,width=6cm}}
\caption{Distribution of events versus radius cubed.}
\end{center}
\end{figure}

Figure 7 shows the distribution of events as a function of (radius/600cm)$^3$, 
for a high-energy threshold.  The radius of the AV is 600 cm, so the heavy water 
volume corresponds to values less than 1. It is apparent that there is a clear 
excess of events in this region, indicating the substantial contribution from 
the CC reaction on deuterium.

\begin{figure}[!ht]
\begin{center}
\mbox{\epsfig{file=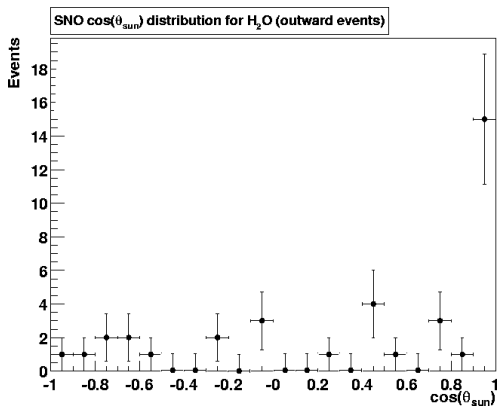,width=6cm}}
\caption{Distribution of events in H$_2$O versus cos$ \theta_{sun}$.}
\end{center}
\end{figure}

Figure 8 shows the distribution of events with a high energy threshold for a 
region in the light water outside the AV. Events have been selected to remove 
inward-coming gamma rays. The peak from ES events is apparent with a relatively 
small background.

\section{RADIOACTIVE BACKGROUNDS}
Radioactive backgrounds that contribute to the Cerenkov light in the detector 
arise from the decay chains of  $^{238}$U and  $^{232}$Th impurities in the 
water and other detector materials. At low energies, the dominant contributions 
come from impurities in the water. These contributions can be measured through 
the radioassay of the light and heavy water. They can also be measured 
independently through observation of the low energy region of the Cerenkov 
spectrum for events reconstructing in the water regions. Sensitive techniques 
have been developed for radioassay of  $^{224}$Ra, $^{226}$Ra and  $^{222}$Rn in the water. The 
measurements for Ra are performed by extracting the Ra on beads coated with 
manganese oxide or on ultrafiltration membranes coated with hydrous titanium 
oxide. After sampling hundreds of tonnes of water, these materials are measured 
for radioactive decay of the Ra with techniques sensitive to tens of atoms. (see 
reference\cite{One} for more details) The $^{222}$Rn is measured by degassing 50 or more tonnes 
of water and collecting the Rn gas with liquid nitrogen-cooled traps. The 
collected gas is then counted with ZnS coated scintillation cells (Lucas cells) 
to observe the alpha decays. 
These techniques have been employed to make very sensitive measurements of the 
water, as shown in Figures 9 and 10.

\begin{figure}[!ht]
\begin{center}
\mbox{\epsfig{file=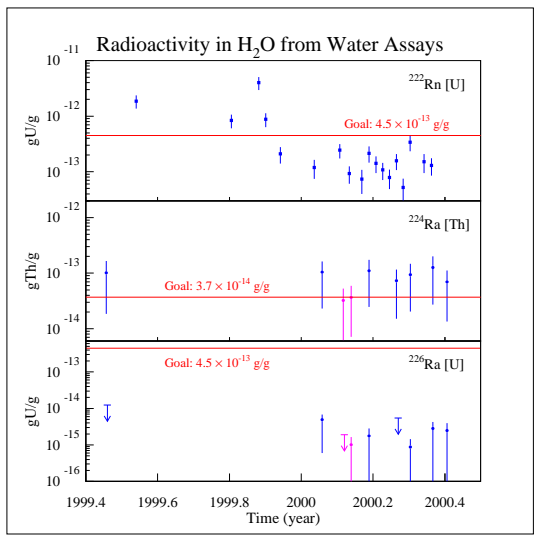,height=7cm}}
\caption{Measurements of Radium from the Th and U chains and Radon from the U 
chain in H$_2$O.}
\end{center}
\end{figure}

The Cerenkov light generated by the Th and U radioactivity can be observed at 
low energies as illustrated in figure 1 and observed in figures 3 and 5. As the 
decay products and sequence are different for the two chains it is also possible to use 
pattern recognition to obtain a statistical separation of the contributions from 
the two chains. Future calibrations will include the use of  proportional 
counters containing Th and U chain sources to provide triggered events to 
calibrate the detector response in this region. However, the data to date, with 
large calibration uncertainties, do agree with the radioassay measurements.  
The light water in the SNO detector is designed to attenuate higher energy gamma 
rays (fission and alpha-induced) from radioactivity in the cavity walls and the 
PMT support structure. High-energy events reconstructed in the light water 
volume outside the AV are found to be predominantly inward going and the numbers 
decrease rapidly as a function of radius. Using calibration data from the $^{16}$N 
source positioned near the PMT's, extrapolations of the number of high energy 
gammas interacting within the D$_2$O volume indicate that fewer than a few percent 
of the events above NHIT = 60 in Figure 5 arise from external high energy 
gammas.
In addition to the contributions to Cerenkov light, the presence of Th and U 
chain elements can produce a background for the NC reaction through the 
photodisintegration of deuterium by 2.6 MeV gammas from the Th chain and 2.4 MeV 
gammas from the U chain. The horizontal lines in Figure 10 individually 
represent contributions to the neutron background in the detector from 
photodisintegration equivalent to 5\% of the signal expected for the NC reaction 
for the neutrino flux\cite{Two} of BP98. As is apparent from the figure, these goals 
have been met for the U chain and are met within a factor of two for the Th 
chain. 

\begin{figure}[!ht]
\begin{center}
\mbox{\epsfig{file=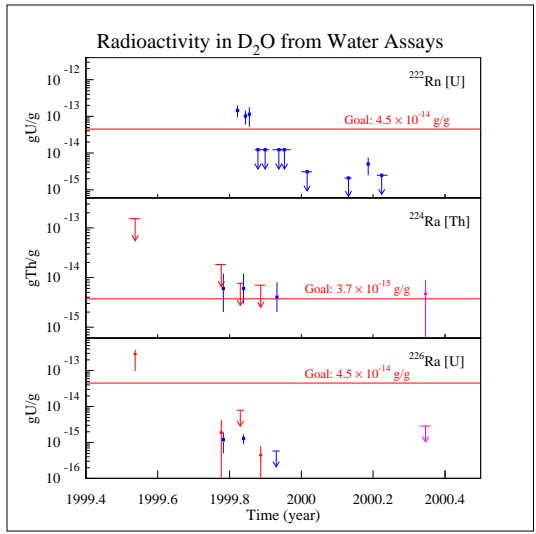,height=7cm}}
\caption{Measurements of Radium from the Th and U chains and Radon from the U 
chain in D$_2$O.}
\end{center}
\end{figure}

\section{CONCLUSIONS}
Based on energy, direction and location  information, the data in the regions of 
interest in Figures 5 to 8 appear to be dominated by $^{8}$B solar neutrino events 
observed with the CC and ES reactions, with very little background. This implies 
that measurements during the pure heavy water phase will provide an accurate 
measurement of the electron neutrino flux via the CC reaction after completion 
of  further calibrations. The measurements of radioactivity  imply that the NC 
measurements can be made with only a small uncertainty from the radioactive 
background.

\section{Acknowledgments}
This research has been financially supported in Canada by the Natural Sciences
 and Engineering Research Council, Industry Canada, National Research
 Council of Canada, Northern Ontario Heritage Fund Corporation and 
the Province of Ontario, in the United States 
by the Department of Energy, and in the 
United Kingdom by the Science and Engineering Research Council and the
 Particle Physics and Astronomy Research Council. Further support was provided by INCO, 
Atomic Energy of Canada Limited (AECL), 
Agra-Monenco, Canatom, Canadian Microelectronics Corporation and Northern
 Telecom. The heavy water has been loaned by 
AECL with the cooperation of Ontario Hydro. 
The provision of INCO of an underground site is greatly appreciated.

$^a$The SNO Collaboration includes: M. G. Boulay, E. Bonvin, M. Chen, F. A. 
Duncan, E. D. Earle, H. C. Evans, G.T. Ewan, R. J. Ford, A. L. Hallin, P. J. 
Harvey, J. D. Hepburn, C. Jillings, H. W. Lee, J. R. Leslie, H. B. Mak, A. B. 
McDonald, W. McLatchie, B. A. Moffat, B.C. Robertson, P. Skensved, B. Sur, {\bf  Queen's 
University, Kingston, Ontario K7L 3N6, Canada.}
% ...

\end{document}